\documentclass[preprint,12pt]{elsarticle}

\usepackage{amssymb}
\usepackage{amsmath}
\usepackage{mathptmx}
\usepackage{graphicx}
\usepackage{booktabs}
\usepackage{xcolor}
\usepackage{lineno}
\usepackage{amsmath}
\usepackage{amssymb}
\usepackage{tikz}
\usepackage{hyperref}
\usepackage{verbatim}
\usepackage{wrapfig}
\usepackage{cleveref}
\usepackage{doi}

\usepackage{hyperref}
\usepackage{cleveref}

\newcommand{\encircled}[1]{\tikz[baseline=(char.base)]{
    \node[shape=circle,draw,inner sep=0.5pt] (char) {#1};}}

\begin{document}

\begin{frontmatter}



\title{GramSeq-DTA: A grammar-based drug-target affinity prediction approach fusing gene expression information} 

\author[label1]{Kusal Debnath} 

\affiliation[label1]{organization={Department of Computer Science, Virginia Commonwealth University},
            addressline={}, 
            city={Richmond},
            postcode={23284}, 
            state={Virginia},
            country={USA}}

\author[label2]{Pratip Rana} 

\affiliation[label2]
{organization={Department of Computer Science, Old Dominion University},
            addressline={}, 
            city={Norfolk},
            postcode={23529},
            state={Virginia},
            country={USA}}

\author[label1]{Preetam Ghosh\corref{cor1}} 
\ead{pghosh@vcu.edu}
\cortext[cor1]{Corresponding author}


\begin{abstract}
Drug-target affinity (DTA) prediction is a critical aspect of drug discovery. The meaningful representation of drugs and targets is crucial for accurate prediction. Using 1D string-based representations for drugs and targets is a common approach that has demonstrated good results in drug-target affinity prediction. However, these approach lacks information on the relative position of the atoms and bonds. To address this limitation, graph-based representations have been used to some extent. However, solely considering the structural aspect of drugs and targets may be insufficient for accurate DTA prediction. Integrating the functional aspect of these drugs at the genetic level can enhance the prediction capability of the models. To fill this gap, we propose GramSeq-DTA, which integrates chemical perturbation information with the structural information of drugs and targets. We applied a Grammar Variational Autoencoder (GVAE) for drug feature extraction and utilized two different approaches for protein feature extraction: Convolutional Neural Network (CNN) and Recurrent Neural Network (RNN). The chemical perturbation data is obtained from the L1000 project, which provides information on the upregulation and downregulation of genes caused by selected drugs. This chemical perturbation information is processed, and a compact dataset is prepared, serving as the functional feature set of the drugs. By integrating the drug, gene, and target features in the model, our approach outperforms the current state-of-the-art DTA prediction models when validated on widely used DTA datasets (BindingDB, Davis, and KIBA). This work provides a novel and practical approach to DTA prediction by merging the structural and functional aspects of biological entities, and it encourages further research in multi-modal DTA prediction. 
\end{abstract}

\begin{graphicalabstract}
\centerline{\includegraphics[width=1\textwidth]{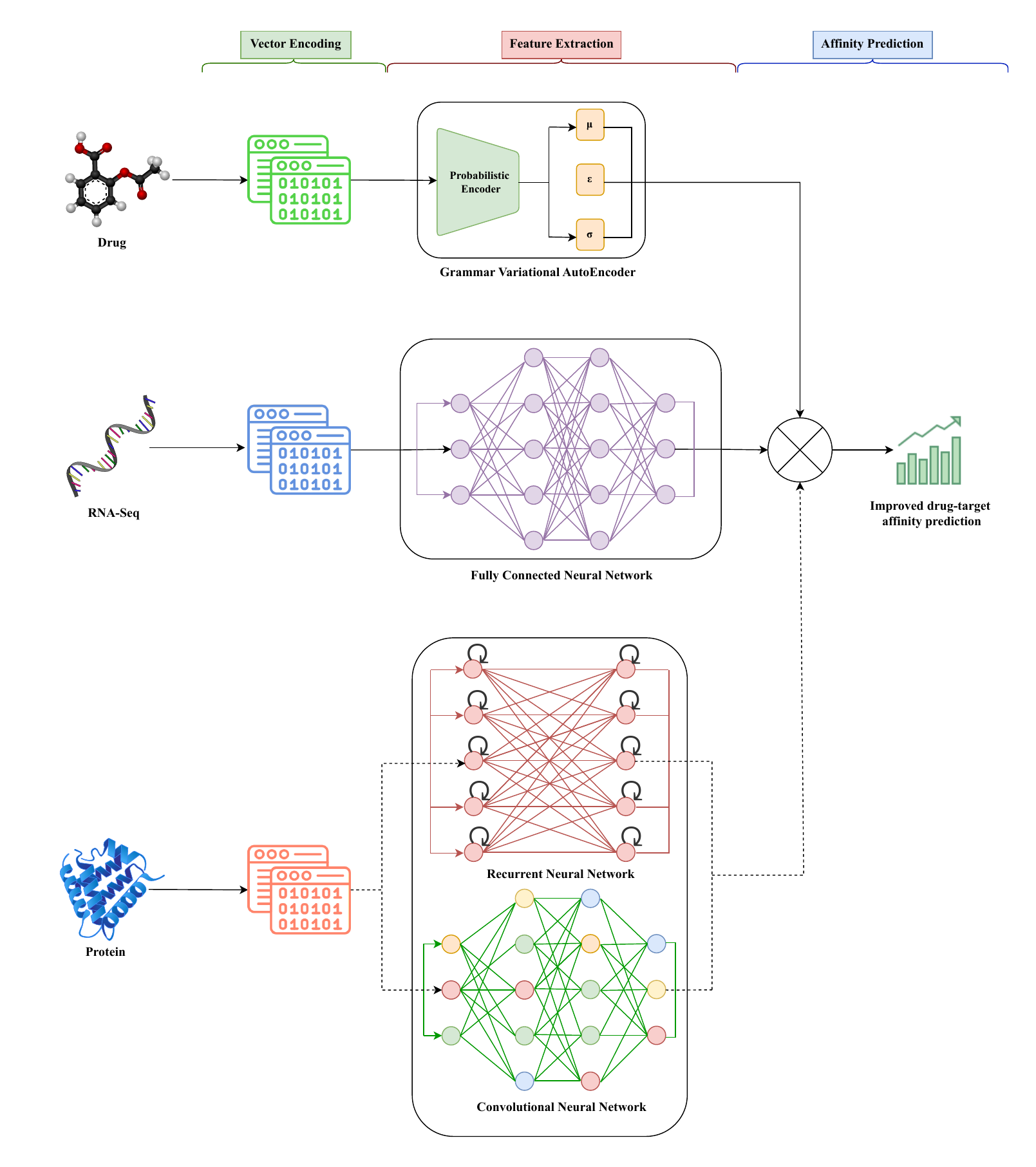}}
\end{graphicalabstract}

\begin{highlights}
\item GramSeq-DTA combines structural and functional representations of drugs and targets, integrating chemical perturbation data with structural information to enhance drug-target affinity (DTA) prediction accuracy.
\item The model uses a Grammar Variational Autoencoder (GVAE) for drug feature extraction besides two different approaches for protein feature extraction, namely Convolutional Neural Network (CNN) and Recurrent Neural Network (RNN).
\item Feature extraction from chemical perturbation information is performed using a Fully Connected Neural Network (FCNN).
\item When validated on widely used DTA datasets (BindingDB, Davis, and KIBA), the proposed approach outperforms current state-of-the-art models by incorporating both genetic and structural data.
\end{highlights}


\end{frontmatter}



\section{Introduction}
\label{sec1}

Drug-target affinity (DTA) prediction provides a foundation for modern drug discovery, bringing various benefits to improve efficiency, reduce costs, and increase success rates. The significance of DTA prediction is well-discussed in current literature, emphasizing its role in accelerating the identification of potential drug candidates and minimizing the risk of failure during clinical trials \cite{mak2023artificial,kim2021comprehensive}. Recent improvements in computational methods \cite{rev1,rev2,rev3,rev4} and the availability of relevant data enhance the accuracy and reliability of DTA predictions\cite{meli2022scoring,moon2024toward}, aiding in the design of efficient therapeutic strategies and effective treatments for many diseases.  

To achieve accurate drug-target affinity (DTA) predictions, the way in which both drugs and targets are represented is a critical determinant of the performance of the models. Proper encoding of these molecular entities is essential for capturing the intricate relationships between their structural and functional properties. Early computational approaches to DTA often relied on simplified representations, such as molecular fingerprints for drugs and amino acid sequences for proteins, to feed machine learning models. Although these methods showed some promise, their inability to fully capture the complexity of molecular interactions limited the predictive power of DTA models\cite{ding2014similarity,pahikkala2015toward,he2017simboost}. As the field evolved, researchers began exploring more sophisticated techniques to better model the structural and chemical properties of drugs and targets, leading to the development of advanced representation methods that significantly improved the predictive accuracy and generalizability of DTA models\cite{ru2021current,chen2024drug,abdul2024comprehensive}.

Ozturk et al.\cite{ozturk2018deepdta} proposed DeepDTA, where they utilized 1D convolutional neural networks (CNNs) to extract high-level representations of protein sequences and 1D SMILES representations of the compounds. Before this approach, most computational methods treated drug-target affinity prediction as a binary classification problem. DeepDTA redefined the problem as a continuum of binding strength values, providing a broader view of drug-target interactions.

Nguyen et al.\cite{nguyen2021graphdta} advanced the field by representing drugs as graphs instead of linear strings and utilized graph neural networks (GNNs) to predict drug-target affinity in their proposed deep learning model called GraphDTA. This approach firmly positions the graph-based representation of drugs as a highly effective and reliable method. Building on this trend, Tran et al.\cite{tran2022deepnc} proposed the Deep Neural Computation (DeepNC) model, which consists of multiple graph neural network algorithms. 

The rise of natural language-based methods in biomedical research has led to further innovations in DTA modeling. Qiu et al.\cite{qiu2024gk} introduced G-K-BertDTA to bridge the gap between the structural and semantic information of molecules. In their approach, drugs were represented as graphs to learn their topological features, and a knowledge-based BERT model was incorporated to obtain the semantic embeddings of the structures, thereby enhancing the feature information. 

Nevertheless, there are some limitations to the above-mentioned approaches. Firstly, the information on the relative positions of the constituent atoms and bonds is often missing in the drug encoding approaches adopted in these models. In addition, the functional aspects of those drugs, which can provide relevant insights into their interaction with targets, were also not incorporated.

To address these limitations, we utilized the encoding approach for drugs known as grammar variational autoencoder (GVAE) proposed by Kusner et al.\cite{kusner2017grammar}. GVAE discusses the parse tree-based encoding of the drug entities, which allows learning from semantic properties and syntactic rules. This approach can learn a more consistent latent space in which entities with nearby representations decode to discrete similar outputs. In addition, to incorporate the functional aspect of those drugs, we integrated the chemical perturbation information from the L1000 project\cite{subramanian2017next}. In the L1000 project, various chemical entities have been used as perturbagens and tested against multiple human cell lines, primarily linked to several types of cancers, to analyze their gene expression profile. We have utilized these gene expression signatures as the functional feature set for the drugs. Thus, the approach taken in this work utilizes structural and functional representation of drugs, which enhances the drug-target affinity prediction and outperforms the current state-of-the-art methods.

The paper provides a thorough overview of the background research in Section 2, along with a detailed explanation of the methodology, dataset preparation, network architecture, and evaluation metrics in Section 3. Section 4 presents the results of the proposed method on commonly used benchmark datasets and compares its performance to the current state-of-the-art DTA prediction models. Finally, Section 5 addresses the limitations of the study and explores opportunities for future research advancements.

\section{Background}
\subsection{Grammar Variational Autencoder}

Gómez-Bombarelli et al. \cite{gomez2018automatic} used Gated Recurrent Units (GRUs) and Deep Convolutional Neural Networks (DCNNs) to develop a generative model for molecular entities based on SMILES strings. This model has the potential to encode and decode molecular entities through a continuous latent space, which aids in the exploration of novel molecules with desirable properties in this space. Nevertheless, one major drawback in using string-based representation for molecular entities is their fragility, i.e., minute alteration in the string can lead to complete deviation from the original molecule, even corresponding to the generation of entirely invalid entities. James et al. \cite{opensmiles} first proposed the concept of constructing grammar for chemical structures. According to this work, every valid discrete entity can be represented as a parse tree from the given grammar. The advantage of generating parse trees compared to texts is that it ensures the complete validity of the generated entities based on grammar. Thus, Kusner et al. \cite{kusner2017grammar} proposed the grammar variational autoencoder (GVAE), which encodes and decodes directly from these parse trees. This approach allows GVAE to learn from syntactic rules as well as to learn semantic properties. Along with its ability to generate valid outputs, this approach can also learn a more coherent latent space in which entities with nearby representations decode to discrete similar outputs. 

\subsubsection{Context-free Grammar}
A context-free grammar (CFG) is conventionally defined as a 4-tuple \emph{G = (V, \( \Sigma \), R, S)},
where \emph{V} represents a finite set of non-terminal symbols; \( \Sigma \) represents a finite set of terminal symbols, disjoint from V; \emph{R} represents a finite set of production rules; \emph{S} is a unique non-terminal referred to as the start symbol; \emph{G} represents the grammar that describes a set of trees that can be formed by applying rules in \emph{R} to leaf nodes until all leaf nodes become terminal symbols in \( \Sigma \).

The rules \emph{R} are technically defined as \(\alpha \to \beta \text{ for } \alpha \in V \text{ and } \beta \in (V \cup \Sigma \ast)\), \( \ast \) denoting the Kleene closure. Practically, these rules are portrayed as a collection of mappings from a solitary non-terminal on the left-hand side in \emph{V} to a sequence of symbols that can be either terminal or non-terminal by definition. These mappings can be seen as a rule for rewriting. 

When a production rule is applied to a non-terminal symbol, it creates a tree structure where the symbols on the right-hand side of the production rule become child nodes for the left-hand side parent. These trees extend from each non-terminal symbol in \emph{V}. The language of \emph{G} is the set of all sequences of terminal symbols that can be generated by traversing the leaf nodes of a tree from left to right. A parse tree is a tree with its root at \emph{S} and a sequence of terminal symbols as its leaf nodes. The prevalence of context-free languages in computer science is attributed, in part, to the existence of practical parsing algorithms.

\subsubsection{Syntactic vs. Semantic Validity}
A crucial aspect of grammar-based encoding is that the encoded molecules are syntactically valid, but the semantic validity of these molecules is a matter of discussion. There are some reasons for this phenomenon - a) Some molecules produced by the grammar may be unstable or chemically invalid; for example, a carbon atom cannot make bonds with more than four atoms in a molecule as it has a valency of 4. Nevertheless, the grammar can produce this kind of molecule; b) Assignment of ring-bond digits in SMILES is a non-context-free process. It needs to keep track of the order in which rings are encountered, along with the connectivity between the rings, which can only be determined from the local context of the string. For example, in naphthalene (c1ccc2c(c1)cccc2), the outer ring uses the digit `1', and the inner ring uses `2'. The digits are not nested but rather follow a specific order; c) GVAE can still produce an undetermined sequence if there are existing non-terminal symbols on the stack after processing all logit vectors.

\subsection{L1000 Assay}

The L1000 project \cite{subramanian2017next} is part of the Library of Integrated Network-Based Cellular Signatures (LINCS) program funded by the National Institutes of Health (NIH). This program aims to catalog and analyze cellular responses to various perturbations to understand how these perturbations modulate cellular functions. This project efficiently manages chemical perturbagen data using a structured method, which encircles data generation, processing, and analysis. This project uses various chemical compounds, including FDA-approved drugs, experimental drugs, natural compounds, and other bioactive molecules as perturbagens. Perturbagens selection mainly involves possible connections of these chemicals to numerous biological pathways and disease cross-talks. Various human cell lines, majorly associated with several types of cancers, are chosen to ensure diversity in the biological responses. The L1000 data thus can be used to identify potential new uses for existing drugs or to discover new candidate drugs. Moreover, novel hypotheses can be made that correspond to the possible effects of the new compounds by performing comparative analyses of the gene expression signatures of known drugs. The data also aids in understanding the underlying molecular mechanism of the diseases by showcasing the effect of different compounds in the alteration of gene expression related to disease pathways.

\section{Methodology}


\subsection{Datasets}
\subsubsection{Benchmark Datasets}
In this study, three datasets are used for the benchmarking purposes: BindingDB \cite{liu2007bindingdb}, Davis \cite{davis2011comprehensive}, and KIBA \cite{tang2014making}. The drug-target affinity dataset in the BindingDB database contains experimental binding affinities between small molecules and protein targets and supplementary information on the entities (e.g., ID, Structure, etc.). In the Davis dataset, the targets are kinase proteins, and the drugs are the small molecules (inhibitors) targeting those kinases. Similar to Davis, the KIBA dataset also focuses on kinase proteins and their corresponding inhibitor drugs, but it contains a more significant number of instances than Davis. 
\par
In these datasets, the drugs are represented as SMILES strings in these datasets, and the target proteins are represented as amino acid sequences. For BindingDB and Davis datasets, the labels are the \( K_d \) (Dissociation constant) values, which indicate the extent of the interactions between each drug-target pair. Meanwhile, a unified KIBA score is used as a label for the KIBA dataset, combining \( K_d \), \( K_i \) (Inhibition Constant), and \( IC_{50} \) (Half Maximal Inhibitory Coefficient) values for corresponding drug-target pairs. The labels are converted into logarithmic form which helps improve the performance of the model in regression tasks.
\subsubsection{L1000 Chemical Perturbation Dataset}
\begin{figure}[t]
\centerline{\includegraphics[width=1\textwidth]{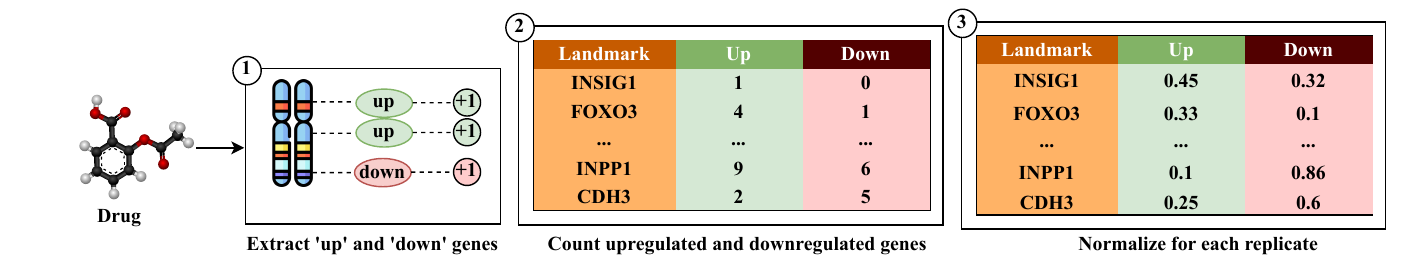}}
\caption{\textbf{Preparation of the gene expression dataset}. Gene expression information analyzed on 978 landmark genes for the selected drugs is extracted from the L1000 chemical perturbation data. After considering all the biological replicates of the perturbation analysis, a gene regulation matrix is created for both upregulated and downregulated genes.}
\label{l1000_prep}
\end{figure}

The chemical perturbation data available in the L1000 project is documented in raw format. Therefore, the data needs to be processed accordingly for use. The detailed process of preparing the L1000 dataset from the raw data is discussed below:

\encircled{1} The L1000 chemical perturbation data file is loaded where each perturbagen has multiple replicates based on dosage concentration, and each replicate has two lists of associated genes - one for upregulated and the other for downregulated. \encircled{2} The analysis of the dosage concentration distribution among the replicates shows that samples with a concentration of \( 10~\mu\text{M} \) are the most common. Therefore, for standardization purposes, samples with a concentration of \( 10~\mu\text{M} \) are selected for further analysis, while the others are excluded. \encircled{3} Each unique perturbagen is mapped into corresponding SMILES representation, which is important for downstream molecular modeling. \encircled{4} For each perturbagen \( p_i \), the gene regulatory information is represented as a vector of `up' and `down' regulation values across 978 landmark genes, and the number of times a gene is upregulated or downregulated is counted and normalized by the number of replicates. Let \(x_{ij}^{up}\) represent whether gene \(j\) is upregulated for perturbagen \(p_i\), and \(x_{ij}^{down}\) represent the same for downregulation. The final regulatory vector for each perturbagen is computed as:
\begin{equation}
\mathbf{v}_i = \frac{1}{\text{count}(p_i)} \left( \sum_{j=1}^{m} x_{ij}^{\text{up}}, \sum_{j=1}^{m} x_{ij}^{\text{down}} \right)
\end{equation}

where \(m = 978\) is the number of landmark genes, and \(\text{count}(p_i)\) is the number of times perturbagen \(p_i\) appears in the dataset. A representative illustration is shown in Figure \ref{l1000_prep}. \encircled{5} Finally, the vector for each perturbagen is stacked to get the final matrix:
\begin{equation}
    \mathbf{V} = \{ \mathbf{v}_i \}_{i=1}^{k}, \quad \mathbf{v}_i \in \mathbb{R}^{978 \times 2}
\end{equation}

where \(\mathbf{V}\) is of shape \(k \times 978 \times 2\), and \(k\) is the number of unique perturbagens. The processed dataset is an important contribution of this work that can aid future research in RNA-Seq data integration and analysis.

Not all the perturbagens mentioned in the L1000 dataset are entirely present in the benchmark datasets. Therefore, the datasets are processed accordingly, and only those drugs whose corresponding regulatory vector is present in the L1000 dataset are selected. The processing of the datasets resulted in a decrease in the number of total interactions. The summary of all the original and processed benchmark datasets is discussed in Table \ref{tab:dataset_summary}.
\begin{table}[t]
    \centering
    \begin{tabular}{lccc}
    \toprule
        Dataset & Compounds & Proteins & Interactions\\
        \midrule
        \textbf{Original}&&&\\
        \quad BindingDB & 22,381 & 1,860 & 91,751\\
        \quad Davis & 68 & 379 & 30,056\\
        \quad KIBA & 2,068 & 229 & 118,254\\
        \textbf{Processed}&&&\\
        \quad BindingDB & 444 & 754 & 18,567\\
        \quad Davis & 11 & 379 & 4,169\\
        \quad KIBA & 12 & 194 & 538\\
    \bottomrule
    \end{tabular}
    \caption{Dataset statistics - number of compounds, proteins and interactions}
    \label{tab:dataset_summary}
\end{table}


\subsection{Network Architecture}
\begin{figure}[t]
    \centering
    \includegraphics[width=0.5\textwidth]{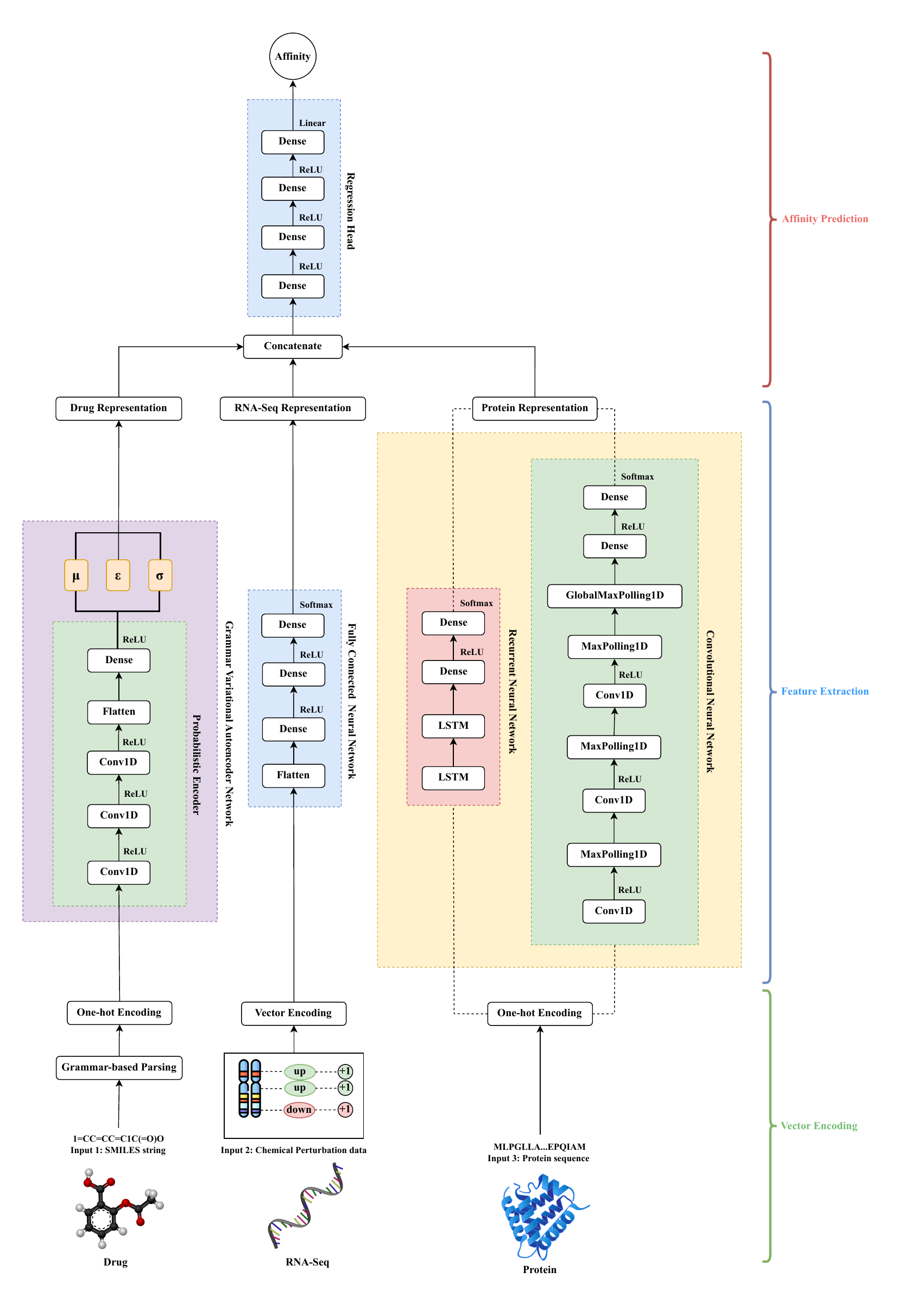}
    \caption{\textbf{Network architecture of the proposed model}. The encoded drug information is passed through an GVAE layer, the RNA-Seq information is passed through an FCNN, while the encoded protein information is passed through a series of LSTM layers and 1D CNN layers. Learned representations are concatenated and passed through a FCNN acting as a regression head to predict the affinity.}
    \label{fig:network_architecture}
\end{figure}

The complete network consists of 3 main parts - a) Drug encoder, b) RNA-Seq encoder, and c) Protein encoder. Figure \ref{fig:network_architecture} shows the schematic diagram of the complete network architecture.

\subsubsection{Drug Encoder}
For this work, we utilized a pre-trained GVAE model from the study conducted by Zhu et al.\cite{zhu2021prediction} that focuses on deep learning-based drug efficacy prediction from transcriptional profiles. One-hot encoded vectors are generated by parsing the SMILES representations of the drugs using a grammar-tree-based approach and then passed into the encoder network. The detailed process of parsing the SMILES and generating one-hot encoded vectors is discussed below:

\encircled{1} SMILES representations are converted into a collection of tokens using a tokenizer.
\encircled{2} The tokenized sequence is then parsed using a grammar adopted from the work of Kusner et al. \cite{kusner2017grammar}. This yields a sequence of production rules:
\begin{equation}
G(\tau(S)) = P = \{P_1, P_2, \dots, P_q\}
\end{equation}

where \(G\) is the grammar, \(\tau(S)\) is the tokenized sequence and \( P = \{P_1, P_2, \dots, P_q\}\) is the sequence of production rules. Each production rule is then mapped to an index \(I_i\) in a predefined list of rules. 
\encircled{3} A zero matrix is initialized, denoting the vector to be populated by one-hot encoding:
\begin{equation}
O_{j,I_j} = 1, \quad \forall j \in \{1, 2, \dots, \min(M, q)\}, \quad I_j \in \{1, 2, \dots, N-1\}
\end{equation}

where \(O\) is the encoded one-hot vector of shape \(M \times N\), \(M\) is the maximum length of sequences, N is the total number of production rules, and \(q\) is the number of productions. If q is smaller than M, the rest of the matrix is padded with an indicator for "end of sequence":
\begin{equation}
O_{j,N-1} = 1, \quad \forall j \in \{q+1, q+2, \dots, M\}
\end{equation}

\begin{figure}[t]
    \centering
    \includegraphics[width=1\textwidth]{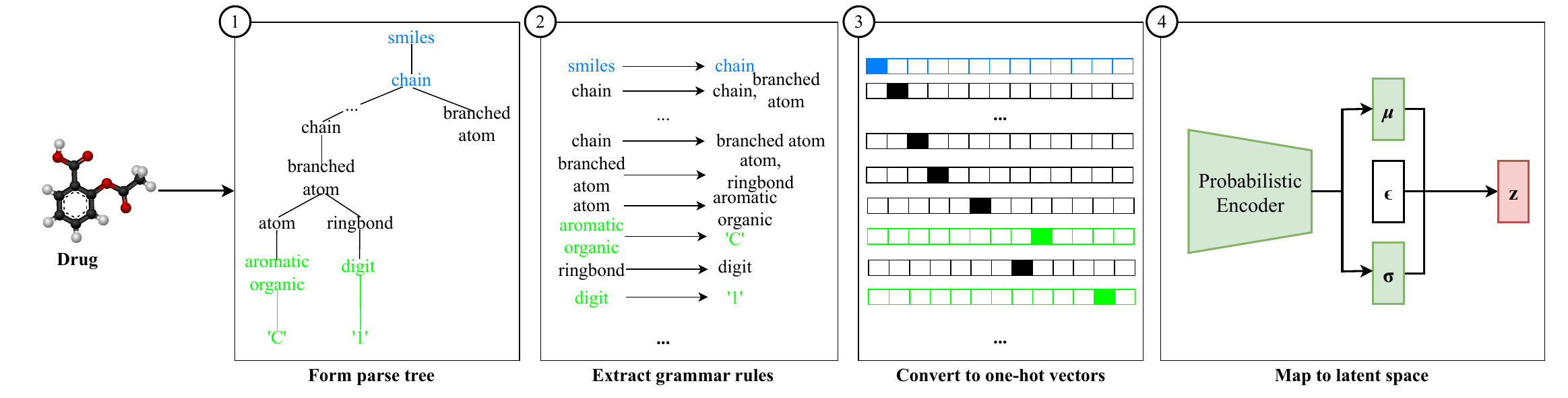}
    \caption{\textbf{Encoding of drug SMILES structures}. A parse tree is constructed based on the structural components of SMILES representations. Grammar rules are extracted from the parsed trees. SMILES representations are then converted into one-hot vectors. Finally, the one-hot vectors are transformed into corresponding latent space representations using an encoder network.}
    \label{fig:gram_enc}
\end{figure}

In this work, the values of \(M\) and \(N\) are 277 and 76, respectively. The generated one-hot vectors for each SMILES representations are then passed into the encoder network. The schematic diagram of the overall process is given in Figure \ref{fig:gram_enc}.

\subsubsection{RNA-Seq Encoder}
A fully connected neural network (FCNN) is used for the extraction of meaningful features from the high-dimensional RNA-Seq data. As discussed in the L1000 chemical perturbation dataset preparation, the resulting dataset is of shape \(k \times 978 \times 2\), where \(k\) is the number of unique perturbagens, 978 is the number of landmark genes and 2 indicates the number of columns representing upregulated and downregulated genes. When these vectors are passed through a dense neural network, it learns a condensed and abstract representation of how each perturbagen affects gene expression. 

\subsubsection{Protein Encoder}
Feature extraction from amino acid sequences is achieved using two different types of neural networks: Convolutional Neural Network (CNN) and Recurrent Neural Network (RNN). CNNs are able to identify local motifs and patterns within a sequence by using sliding windows of filters to capture neighboring amino acids. On the other hand, RNNs are capable of extracting long-range dependencies and sequential relationships between amino acids by retaining information from previous positions in the sequence in a step-by-step manner. To encode the sequences, a dictionary of all possible amino acid sequences in the proteins is created. One-hot encoding of a given protein is carried out based on the presence of a particular amino acid in that protein. For standardization, the maximum length of a protein is limited to 1000 sequences. Encoding of all the proteins results in the creation of a vector of \(p \times 26 \times 1000\), where \(p\) is the number of unique proteins and 26 is the length of the amino acid dictionary.

\subsection{Training Settings}
The training process is set to run for 500 epochs with an adaptive learning rate that starts at 0.001 using the Adam optimizer. The higher learning rate value is chosen to ensure that the training process does not significantly impact the pre-trained weights in the GVAE model. A batch size of 256 has been found to yield the best results, maintaining a balance between memory usage and convergence speed. The summary of the overall network architecture and training hyperparameters are discussed in Table \ref{tab:table_parameters}.

\begin{table}[t]
    \centering
    \begin{tabular}{lll}
        \toprule
        Parameters&Value \\
        \midrule
        \textbf{Drug Encoding}&\\
        \quad GVAE Encoder Filter Sizes&9, 9, 10\\
        \quad GVAE Encoder Kernel Sizes&9, 9, 11\\
        \quad GVAE Latent Space Dimension&56\\
        \textbf{RNA-Seq Encoding}&\\
        \quad Dense Layers&2\\
        \textbf{Protein Encoding}&\\
        \quad CNN Filter Sizes&32, 64, 96\\
        \quad CNN Kernel Sizes&4, 8, 12\\
        \quad RNN LSTM Layers&2\\
        \textbf{Regression Head}&\\
        \quad Dense Layers&3\\
        \textbf{Training}&\\
        \quad Epochs&500\\
        \quad Learning Rate&0.001\\
        \quad Batch Size&256\\
        \quad Optimizer&Adam\\
        \bottomrule
    \end{tabular}
    \caption{Summary of Network Architecture and Training Hyperparameters}
    \label{tab:table_parameters}
\end{table}

\subsection{Evaluation Metrics}
Evaluating deep learning models involves various metrics that capture different aspects of performance. Mean Squared Error (MSE) measures the average squared difference between actual and predicted values, highlighting prediction accuracy.
We also used the Concordance Index (C-Index), which is a preferred metric for survival analysis, to evaluate the consistency between predicted risk scores and actual outcomes. Together, these metrics provide a robust framework for model evaluation.

\subsubsection{Mean Squared Error (MSE)}
Mean Squared Error (MSE) is a common loss function used for regression tasks. It measures the average of the squares of the errors, which are the differences between the predicted and actual values. Mathematically, it is defined as:
\begin{equation}
\text{MSE} = \frac{1}{n} \sum_{i=1}^{n} (y_i - \hat{y}_i)^2
\end{equation}
where \( y_i \) is the actual value, \( \hat{y}_i \) is the predicted value, and \( n \) is the number of data points.


\subsubsection{Concordance Index (C-Index)}
The Concordance Index (C-Index) is a metric used primarily in survival analysis to evaluate the predictive accuracy of risk scores. It assesses the degree of concordance between the predicted and actual ordering of event times. The C-Index is calculated as:
\begin{equation}
C = \frac{\text{Number of concordant pairs}}{\text{Number of possible evaluation pairs}}
\end{equation}
A pair is considered concordant if the predicted and actual orderings of two instances are consistent.

\section{Results \& Discussions}


In this section, we will discuss the performance of the GramSeq-DTA model in detail. The discussion can be divided into the following parts: a) Benchmarking the performance of GramSeq-DTA with integrated RNA-Seq information against the baseline models, b) Advantage of integrating RNA-Seq information to the model, and c) Performance comparison of the proposed model on original and processed datasets
\par
\subsection{Benchmarking against baseline models}
In order to validate our findings, we conducted a comprehensive performance comparison of GramSeq-DTA, which now includes integrated RNA-Seq information. We compared it against several well-established baseline models: DeepDTA, GraphDTA, DeepNC, and G-K-BertDTA. 
We used MSE and CI values to assess performance. The model performance is compared across three benchmark datasets - BindingDB, Davis, and KIBA. DeepDTA employs a deep learning framework to capture the complex features of drug-target interactions using convolutional neural networks. GraphDTA, on the other hand, leverages graph neural networks to represent molecular structures as graphs, enabling it to better capture the topological properties of molecules. DeepNC uses a neural collaborative filtering approach to model interactions, focusing on latent feature extraction. Lastly, G-K-BertDTA integrates graph-based representations with BERT-like architectures to enhance contextual understanding of molecular relationships. Our extensive evaluation, conducted on processed benchmark datasets, showed that GramSeq-DTA consistently outperformed its counterparts in terms of Concordance Index (CI) values, a widely accepted metric for evaluating predictive performance in drug-target affinity modeling. 
In the BindingDB dataset, GramSeq-DTA outperforms G-K-BertDTA by 1.32\% in terms of CI value. Similarly, on the Davis dataset, GramSeq-DTA shows a 0.89\% advantage over DeepNC. On the KIBA dataset, GramSeq-DTA exceeds G-K-BertDTA by 2.75\%.
Importantly, the integration of RNA-Seq data into GramSeq-DTA provided valuable insights into gene expression patterns, contributing to the improved accuracy of the models in predicting drug-target interactions. Detailed results of this comparison can be found in Tables \ref{tab:gramseq_baselines_bindingdb}, \ref{tab:gramseq_baselines_davis}, and \ref{tab:gramseq_baselines_kiba}, where the enhanced GramSeq-DTA model demonstrates its robust performance, setting a new standard in the field.

\begin{table*}[t]
    \centering
    \begin{tabular}{llllll}
    \toprule
        Model&Drug&RNA-Seq&Protein&MSE&CI\\
        &Encoder&Encoder&Encoder&&\\
        \midrule
        DeepDTA&CNN&-&CNN&0.384&0.821\\
        GraphDTA&GINConvNet&-&CNN&0.355&0.819\\
        &GCNNet&-&CNN&0.397&0.786\\
        &GATNet&-&CNN&0.512&0.757\\
        &GAT\_GCN&-&CNN&0.384&0.806\\
        DeepNC&GENConv&-&CNN&0.367&0.828\\
        G-K-BertDTA&GINConvNet + Embeddings&-&CNN&\textbf{0.325}&0.832\\
        GramSeq-DTA&GVAE&FCNN&CNN&0.365&\textbf{0.843}\\
        &GVAE&FCNN&RNN&0.355&0.832\\
    \bottomrule
    \end{tabular}
    \caption{Performance comparison of GramSeq-DTA with baseline models on the processed BindingDB dataset.}
    \label{tab:gramseq_baselines_bindingdb}
\end{table*}

\begin{table*}[t]
    \centering
    \begin{tabular}{llllll}
    \toprule
        Model&Drug&RNA-Seq&Protein&MSE&CI\\
        &Encoder&Encoder&Encoder&&\\
        \midrule
        DeepDTA&CNN&-&CNN&0.219&0.779\\
        GraphDTA&GINConvNet&-&CNN&0.187&0.771\\
        &GCNNet&-&CNN&0.214&0.732\\
        &GATNet&-&CNN&0.241&0.713\\
        &GAT\_GCN&-&CNN&0.227&0.753\\
        DeepNC&GENConv&-&CNN&0.198&0.789\\
        G-K-BertDTA&GINConvNet + Embeddings&-&CNN&\textbf{0.169}&0.778\\
        GramSeq-DTA&GVAE&FCNN&CNN&0.293&\textbf{0.796}\\
        &GVAE&FCNN&RNN&0.261&\textbf{0.796}\\
    \bottomrule
    \end{tabular}
    \caption{Performance comparison of GramSeq-DTA with baseline models on the processed Davis dataset.}
    \label{tab:gramseq_baselines_davis}
\end{table*}

\begin{table*}[t]
    \centering
    \begin{tabular}{llllll}
    \toprule
        Model&Drug&RNA-Seq&Protein&MSE&CI\\
        &Encoder&Encoder&Encoder&&\\
        \midrule
        DeepDTA&CNN&-&CNN&0.877&0.609\\
        GraphDTA&GINConvNet&-&CNN&1.061&0.628\\
        &GCNNet&-&CNN&0.903&0.631\\
        &GATNet&-&CNN&0.957&0.609\\
        &GAT\_GCN&-&CNN&0.831&0.671\\
        DeepNC&GENConv&-&CNN&0.769&0.648\\
        G-K-BertDTA&GINConvNet + Embeddings&-&CNN&\textbf{0.693}&0.689\\
        GramSeq-DTA&GVAE&FCNN&CNN&0.843&\textbf{0.708}\\
        &GVAE&FCNN&RNN&1.269&0.688\\
    \bottomrule
    \end{tabular}
    \caption{Performance comparison of GramSeq-DTA with baseline models on the processed KIBA dataset.}
    \label{tab:gramseq_baselines_kiba}
\end{table*}

\subsection{Advantage of integrating RNA-Seq information}
Table \ref{tab:gramseq_rnaseq_bindingdb} indicates that when validated on the processed BindingDB dataset, GramSeq-DTA performs better, with a CI value of 0.843 when integrating RNA-Seq information compared to not integrating RNA-Seq information. Similar results are evident in Table \ref{tab:gramseq_rnaseq_davis} and Table \ref{tab:gramseq_rnaseq_kiba}, where validation is performed on the Davis and KIBA datasets, respectively. CI values of 0.796 and 0.708 are observed for the processed Davis and KIBA datasets, respectively, when RNA-Seq information is integrated. These observations prove that integrating RNA-Seq information with corresponding drug and target structural information can enhance the drug-target affinity prediction ability of the model.

\begin{table*}[t]
    \centering
    \begin{tabular}{llllll}
    \toprule
        Model&Drug&RNA-Seq&Protein&MSE&CI\\
        &Encoder&Encoder&Encoder&&\\
        \midrule
        GramSeq-DTA&GVAE&-&CNN&0.495&0.746\\
        &GVAE&-&RNN&0.495&0.754\\
        &GVAE&FCNN&CNN&0.365&\textbf{0.843}\\
        &GVAE&FCNN&RNN&\textbf{0.355}&0.832\\
    \bottomrule
    \end{tabular}
    \caption{Performance comparison of GramSeq-DTA with and without RNA-Seq information integration on the processed BindingDB dataset.}
    \label{tab:gramseq_rnaseq_bindingdb}
\end{table*}

\begin{table*}[t]
    \centering
    \begin{tabular}{llllll}
    \toprule
        Model&Drug&RNA-Seq&Protein&MSE&CI\\
        &Encoder&Encoder&Encoder&&\\
        \midrule
        GramSeq-DTA&GVAE&-&CNN&0.277&0.705\\
        &GVAE&-&RNN&0.311&0.716\\
        &GVAE&FCNN&CNN&0.293&\textbf{0.796}\\
        &GVAE&FCNN&RNN&\textbf{0.261}&\textbf{0.796}\\
    \bottomrule
    \end{tabular}
    \caption{Performance comparison of GramSeq-DTA with and without RNA-Seq information integration on the processed Davis dataset.}
    \label{tab:gramseq_rnaseq_davis}
\end{table*}

\begin{table*}[t]
    \centering
    \begin{tabular}{llllll}
    \toprule
        Model&Drug&RNA-Seq&Protein&MSE&CI\\
        &Encoder&Encoder&Encoder&&\\
        \midrule
        GramSeq-DTA&GVAE&-&CNN&1.011&0.653\\
        &GVAE&-&RNN&0.876&0.618\\
        &GVAE&FCNN&CNN&\textbf{0.843}&\textbf{0.708}\\
        &GVAE&FCNN&RNN&1.269&0.688\\
    \bottomrule
    \end{tabular}
    \caption{Performance comparison of GramSeq-DTA with and without RNA-Seq information integration on the processed KIBA dataset.}
    \label{tab:gramseq_rnaseq_kiba}
\end{table*}

\subsection{Performance on original and processed data}
Table \ref{tab:gramseq_benchmarks_ori} presents the comparative evaluation of the performance of the proposed model on the original benchmark datasets and the processed benchmark datasets. The results of the model integration with RNA-Seq information are shown for the processed datasets. As shown in Table \ref{tab:dataset_summary}, the number of interactions between the original and the processed datasets differs. Despite losing approximately 80\% of data in processing, our model performs better on the processed BindingDB dataset, with a best CI value of 0.843, while for the original BindingDB dataset, the optimum CI value was 0.818. The results on the original and processed Davis dataset are also competitive. During processing the Davis dataset, we lost around 84\% of data. Our model indicates a CI value of 0.809 for the original Davis dataset, while for the processed Davis dataset, the CI value is 0.796. Data loss during the processing of the KIBA dataset is 99.5\%, which is the highest value among all three datasets. For the KIBA dataset, there is a significant difference in CI values (original: 0.823, processed: 0.708) for original and processed datasets. The performance difference for the KIBA dataset can be caused by excessive data loss while processing the original dataset. Based on the performance of the other two process datasets (BindingDB and Davis), with more data available for the processed KIBA dataset, there is a high possibility of getting better performance.
\begin{table*}[t]
    \centering
    \begin{tabular}{lllllll}
    \toprule
        Dataset&Model&Drug&RNA-Seq&Protein&MSE&CI\\
        &&Encoder&Encoder&Encoder&&\\
        \midrule
        BindingDB&&&&&\\
        \midrule
        \quad Original&GramSeq-DTA&GVAE&-&CNN&1.029&0.818\\
        &&GVAE&-&RNN&1.061&0.812\\
        \quad Processed&GramSeq-DTA&GVAE&FCNN&CNN&0.365&\textbf{0.843}\\
        &&GVAE&FCNN&RNN&\textbf{0.355}&0.832\\
        \midrule
        Davis&&&&&\\
        \midrule
        \quad Original&GramSeq-DTA&GVAE&-&CNN&0.446&0.806\\
        &&GVAE&-&RNN&0.445&\textbf{0.809}\\
        \quad Processed&GramSeq-DTA&GVAE&FCNN&CNN&0.293&0.796\\
        &&GVAE&FCNN&RNN&\textbf{0.261}&0.796\\
        \midrule
        KIBA&&&&&\\
        \midrule
        \quad Original&GramSeq-DTA&GVAE&-&CNN&\textbf{0.272}&\textbf{0.823}\\
        &&GVAE&-&RNN&0.277&\textbf{0.823}\\
        \quad Processed&GramSeq-DTA&GVAE&FCNN&CNN&0.843&0.708\\
        &&GVAE&FCNN&RNN&1.269&0.688\\
    \bottomrule
    \end{tabular}
    \caption{Performance GramSeq-DTA without RNA-Seq information integration on the original benchmark datasets.}
    \label{tab:gramseq_benchmarks_ori}
\end{table*}

\section{Conclusion \& Future Directions}
In this study, we demonstrated that incorporating chemical perturbation information can enhance drug-target affinity prediction. The core contribution of this research is in its transformation of data from a chemical perturbation assay and utilizing it as an extra modality along with drug and protein structural information. This research could guide future work on a better understanding of affinity prediction among biological entities in the absence of three-dimensional structural information of the entities. Nevertheless, a fundamental limitation of this research is that information from a chemical perturbation assay may not be available for every drug in widely used drug-target affinity benchmark datasets. Therefore, having more data from chemical perturbation assays for additional drugs can further enhance the ability of deep learning models to predict affinity. Future studies should investigate different approaches for transforming the chemical perturbation information. Moreover, introducing advanced feature extraction methods from the biological entities can enhance the prediction. In summary, this work underscores the importance of integrating additional data modalities in drug-target affinity prediction. 












\section*{CRediT authorship contribution statement}
\noindent \textbf{Kusal Debnath:} Data Curation, Conceptualization, Methodology, Writing - Original Draft Preparation, Writing - Review \& Editing, Visualization. 
\textbf{Pratip Rana:} Conceptualization, Formal Analysis, Investigation, Writing - Original Draft Preparation, Writing - Review \& Editing.
\textbf{Preetam Ghosh:} Conceptualization, Resources, Writing - Review \& Editing, Project Administration, Supervision, Funding Acquisition.

\section*{Acknowledgment}

\noindent  This work was partially supported by 5R21MH128562-02 (PI: Roberson-Nay), 5R21AA029492-02 (PI: Roberson-Nay), CHRB-2360623 (PI: Das), NSF-2316003 (PI: Cano), VCU Quest (PI: Das) and VCU Breakthroughs (PI: Ghosh) funds awarded to P.G.

\section*{Conflict of Interest}

\noindent  The authors declare that they have no conflict of interest regarding the publication of this paper.

\section*{Author Biographies}

\begin{minipage}{0.3\textwidth}
    \includegraphics[width=\textwidth]{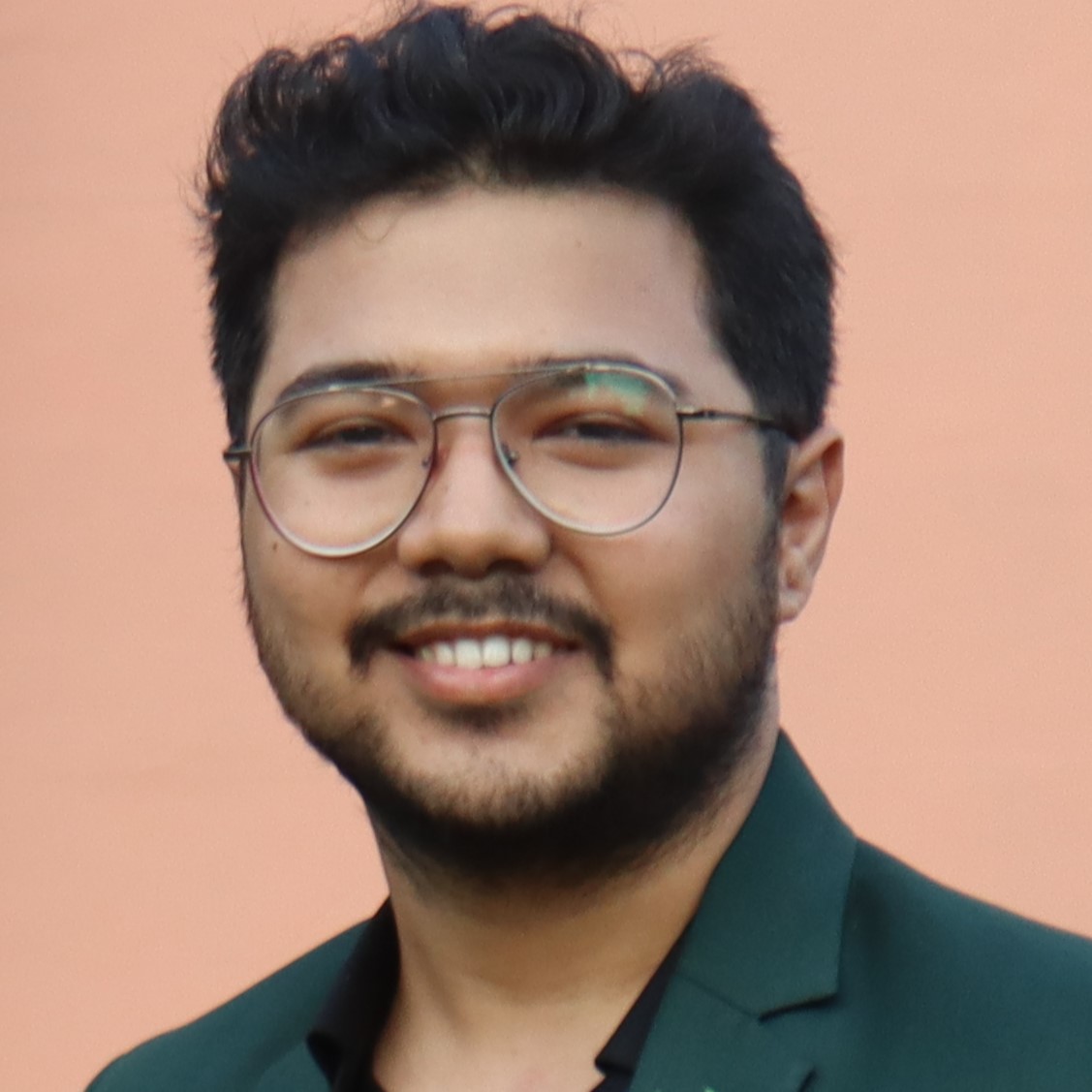} 
\end{minipage}
\begin{minipage}{0.7\textwidth}
    \textbf{Kusal Debnath} is pursuing PhD in Computer Science with a strong focus on AI-driven drug discovery. He obtained his Master of Technology (MTech) degree in Biomedical Engineering from Indian Institute of Technology, Kharagpur, India. His research interests include machine learning applications in bioinformatics and computational biology, and he is passionate about advancing innovations at the intersection of technology and medicine.
\end{minipage}

\vspace{1cm} 

\begin{minipage}{0.3\textwidth}
    \includegraphics[width=\textwidth]{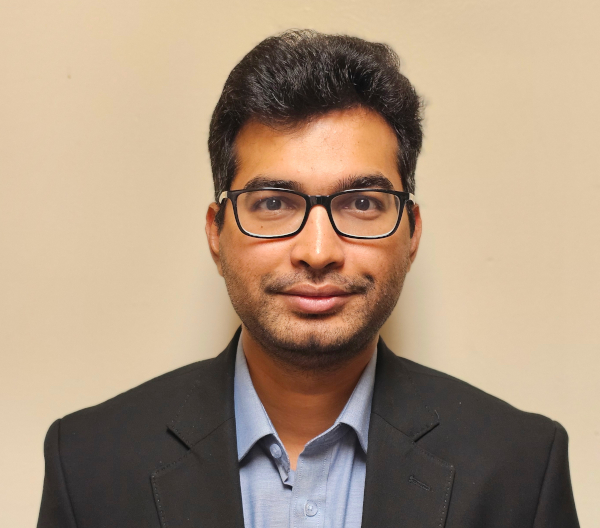} 
\end{minipage}
\begin{minipage}{0.7\textwidth}
    \textbf{Pratip Rana} is an assistant professor in the Department of Computer Science at Old Dominion University. He received a Ph.D. degree in Computer Science from Virginia Commonwealth University, USA, in 2020. His research interests include Machine learning, Computational biology, Complex systems, and Modeling \& simulation.
\end{minipage}

\vspace{1cm} 

\begin{minipage}{0.3\textwidth}
    \includegraphics[width=\textwidth]{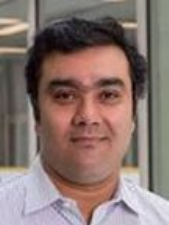} 
\end{minipage}
\begin{minipage}{0.7\textwidth}
    \textbf{Preetam Ghosh} is a Professor in the Department of Computer Science and the director of the Biological Networks Lab at Virginia Commonwealth University. He obtained his MS and Ph.D. degrees in Computer Science \& Engineering from the University of Texas at Arlington and a BS in Computer Science from Jadavpur University, Kolkata India. His research interests include algorithms, stochastic modeling \& simulation, network science and machine learning-related approaches in systems biology and computational epidemiology, and mobile computing-related issues in pervasive grids that have resulted in more than 200 conference and journal articles and several federally funded research projects from NSF, NIH, DoD, and US-VHA. He previously served as the Secretary/Treasurer of ACM SIGBio.
\end{minipage}

\bibliographystyle{unsrt}
\bibliography{refs}

\end{document}